\documentclass[reprint,amsmath,amssymb,aps,showpacs]{revtex4-1} 

\usepackage{graphicx}
\usepackage{dcolumn}
\usepackage{bm}
\usepackage{hyperref}
\usepackage{multirow} %
\usepackage{lineno} 

\begin{document}
\title{Single crystal growth and magnetoresistivity study of topological semimetal CoSi}

\author{D. S. Wu$^{1,2}$}
\author{Z. Y. Mi$^{1,2}$}
\author{Y. J. Li$^{1,2}$}
\author{W. Wu$^{1,2}$}
\author{P. L. Li$^{1,2}$}
\author{Y. T. Song$^{1,2}$}
\author{G. T. Liu$^{1,2,3}$}
\author{G. Li$^{1,2,3}$} 
\email{gli@iphy.ac.cn}
\author{J. L. Luo$^{1,2,3}$}
\email{jlluo@iphy.ac.cn}
\affiliation{
$^1$Institute of Physics and Beijing National Laboratory for Condensed Matter Physics, Chinese Academy of Sciences, Beijing 100190, China\\
$^2$School of Physical Sciences, University of Chinese Academy of Sciences, Beijing 100190, China\\
$^3$Songshan Lake Materials Laboratory, Dongguan, Guangdong 523808, China}
\begin{abstract}
We report single crystal growth of CoSi, which has recently been recognized as a new type of topological semimetal hosting fourfold and sixfold degenerate nodes. The Shubnikov-de Haas quantum oscillation (QO) is observed on our crystals. There are two frequencies originating from almost isotropic bulk electron Fermi surfaces, in accordance with band structure calculations. The effective mass, scattering rate, and QO phase difference of the two frequencies are extracted and discussed.

\end{abstract}

\maketitle
CoSi is a long known material with FeSi-type cubic structure (B20, space group P2$_{1}$3), of which the thermoelectric property and application have been the focus of study\cite{JPSJ.76.093601,Inte.Metal.88.46}, it is generally regarded as a semimetal. With the development of characterizing materials by the topology of their electronic band structure, it has recently been suggested that in crystalline systems, in addition to Dirac, Weyl, and nodal line semimetals, band crossing points (nodes) with three-, six-, or eight-fold degeneracies can be stablized\cite{TPSM_Science}, the low energy fermionic excitations close to the nodes are called "new fermions". For CoSi and isostructural transition metal silicides RhSi, RhGe, and CoGe, the crystal structure is asymmorphic with threefold rotation symmetry and twofold screw symmetry. Theoretical studies\cite{PRL.119.206402,PRL.119.206401} found that with inclusion of spin-orbit-coupling (SOC), close to the Fermi level there only exist bands containing a fourfold degenerate node at the Brillouin zone (BZ) center $\Gamma$, and a sixfold degenerate node at BZ corner R, both are chiral, thus could
serve as a model system in the study of unconventional chiral fermions. Besides, the phonon spectra of these monosilicides host double Weyl points\cite{PRL.120.016401}. Subsequent angle resolved photoemission spectrescopy (ARPES) measurements\cite{PhysRevLett.122.076402,Rao2019,Sanchez2019} have confirmed the overall bulk band structure and the existence of surface Fermi arcs. However, it has been noted that for CoSi preparation of a flat surface for ARPES is at least challenging. Therefore other probes of its bulk electronic structure are needed. At the same time, in the context of searching for feature characteristic of chiral fermions, the physical properties of CoSi are also worth revisiting on high quality single crystalline samples.

In this work, we report single crystal growth by the flux method, different from the Czochralski, Bridgman, or chemical vapor transport techniques previously used. The resulting crystals are of decent quality. To our knowledge, this is the first time SdH is observed in CoSi. The angular dependence of the QO frequencies reveals that they are from the bulk electron Fermi surface centered at R in BZ, agreeing with band calculation and ARPES data. Thus, it is a good starting point for further investigations.

\begin{figure}[htbp]
\includegraphics[width=0.85\columnwidth]{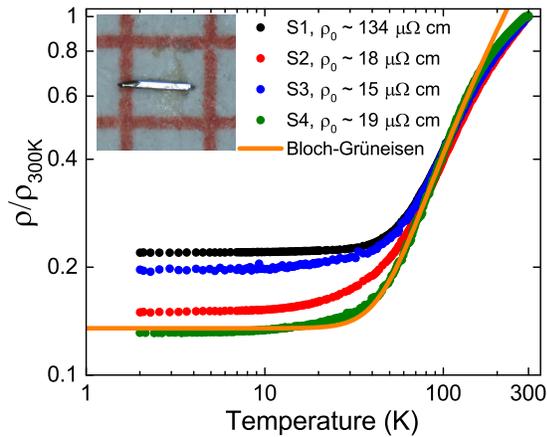}
\caption{\label{fig-RT}(Color online) Normalized temperature-dependent resistivity of several crystals in log-log scale. The corresponding resistivity at 2 K for each sample is labeled in the legend. The solid line represents a simulation by Bloch-Gr$\ddot{\textrm{u}}$neisen formula(see text) for S4. The inset is an optical image of a prismatic rod-shaped single crystal on millimeter scale.
}
\end{figure}

The single crystal growth starts with Co powder (99.999\%, Alfa), Si powder (99.999\%, Alfa), and Sn grains (99.99\%), which were mixed with a molar ratio of 1:1:20, loaded into an alumina crucible, then sealed with a partial Ar pressure inside a silica capsule. The reactant was quickly heated up to 1273 K and maintained there for 40 hours, followed by cooling down in 3 days to 1073 K, at which temperature the flux was separated by centrifuge. The harvested crystals with well-defined surfaces were further washed by dilute HCl to remove possible tin coating. There were mainly two types of shapes, one is thin, prismatic rod, the other is polyhedron close to a cube. To check for chemical composition, we employed energy dispersive X-ray (EDX) spectroscopy equipped on a Hitachi S-4800 scanning electron microscope (SEM). The rod type crystals showed good consistency among different growth batches with the Co:Si atomic ratio close to unit, the Co percentage was ranging from 49.5\% to 50.7\%. However, the polyhedron type crystals showed a large variation in composition, with the Co\% expanding from 39\% to 61\%, indicating formation of other Co-Si binary compounds\cite{Co-Si_phase}. The quality of the rod type crystals were then checked by single crystal x-ray diffraction at room temperature in a Bruker D8 Venture diffractometer using Mo K¦Á radiation, $\lambda$ = 0.71073 \AA. Structure refinement was performed by the program SHELXL-2014/7\cite{SHELX} embedded in the program suite Apex3. Employing the known crystal structure of CoSi, the best refinement had R-indices of R = 0.0334(74) and wR2 = 0.0767(74), respectively. The lattice constant determined is a=4.4376(6) \AA, in accordance with reported values\cite{PRB.86.064433}. In addition, the orientation along the rod is determined as the $\langle011\rangle$ direction of the crystal, an optical image of a typical crystal in shown as inset of Fig.\ref{fig-RT}.

Based on initial composition and crystallographic screening, all the electrical resistivity measurements were performed on rod-type single crystals by standard four-probe method with electrodes along the rod made by silver paste (DuPont 4922N) and platinum wires. Environment of cryogenic temperature and magnetic field was provided by either a 14 Tesla Quantum Design physical property measurement system(PPMS-14), or an Oxford 14 Tesla magnet with top-loading He-3 insert, both equipped with single-axis rotator. An external AC current source and lock-in amplifier were utilized. We note here that there are only a few inconsistent reports\cite{HeatCap72,Narozhnyi2013} on the temperature and magnetic field dependent magnetization of CoSi . Due to our current crystal size and limitation of instrumental resolution, it is planned in our future work.

The temperature dependence of resistivity $\rho(T)$ with $I\parallel\langle011\rangle$ for several samples is shown in Fig.\ref{fig-RT}. It is overall metallic while there is no up-turn at low temperature, against that of sample having nostoichiometric cobalt\cite{Semicon}. The residual resistivity $\rho$(2 K) for 3 out of 4 samples is around 20 $\mu\Omega$ cm, at the low side of reported data\cite{PRB.86.064433,PRB.82.155124,JMMM.177.1371,EPL.120.17002,PhysicaB.244.138}.
The largest residual resistivity ratio [RRR = $\rho$(300 K)/$\rho$(2 K)] appears in Sample S4, RRR$\sim$7.6, at the high side of reported values. Since the residual resistivity and the RRR are commonly used as indicators of crystal quality of metallic systems, we judge that our flux grown crystals have at least reached the same level of quality as crystals grown by other methods. Compared to other known Dirac and Weyl semimetals, e.g., Cd$_{3}$As$_{2}$, for which residual resistivity $\sim$ 20 $n\Omega$ cm, and RRR$\sim$4000 has been reported\cite{Cd3As2_MR}, the corresponding values for CoSi are moderate. Simply considering that the residual resistivity is proportional to the product of carrier density and carrier mobility, while the carrier density of Cd$_{3}$As$_{2}$ is lower than that of CoSi($\sim$10$^{19}$ vs. $\sim$10$^{20}$ cm$^{-3}$), the carrier mobility of CoSi should be lower by orders of magnitude. Whether it is intrinsic property of CoSi or that samples at current stage still have high level of impurity scattering requires further investigation. A simulation of $\rho(T)$ is also tested by the Bloch-Gr$\ddot{\textrm{u}}$neisen (BG) formula\cite{Cvijovi2011}:
\begin{equation}
\rho(T) = \rho_{0} + C(\frac{T}{\Theta_{R}})^{n}\int^{\frac{\Theta_{R}}{T}}_{0}\frac{t^{n}}{(e^{t}-1)(1-e^{-t})}dt
\end{equation}
in which $\Theta_{R}$ is a characteristic temperature that is usually equal to the Debye temperature $\Theta_{D}$, and n is an integer depending on the scattering mechanism, n = 5 for electron-phonon scattering. As shown in Fig.\ref{fig-RT}, for sample S4 BG simulation with n fixed at 5 could only closely follow the temperature dependence below 170 K, while the derived $\Theta_{R}$ is 300 K, only about 60\% of $\Theta_{D}$ obtained by heat capacity measurement\cite{HeatCap72}. Therefore, it is clear that an one channel electron-phonon scattering mechanism could not explain the $\rho(T)$ behavior of CoSi. Other causes, including an non-constant $\Theta_{D}$, multiple conducting channels, and electron-electron scattering, may also need to be considered.

\begin{figure}[htbp]
\includegraphics[width=0.95\columnwidth]{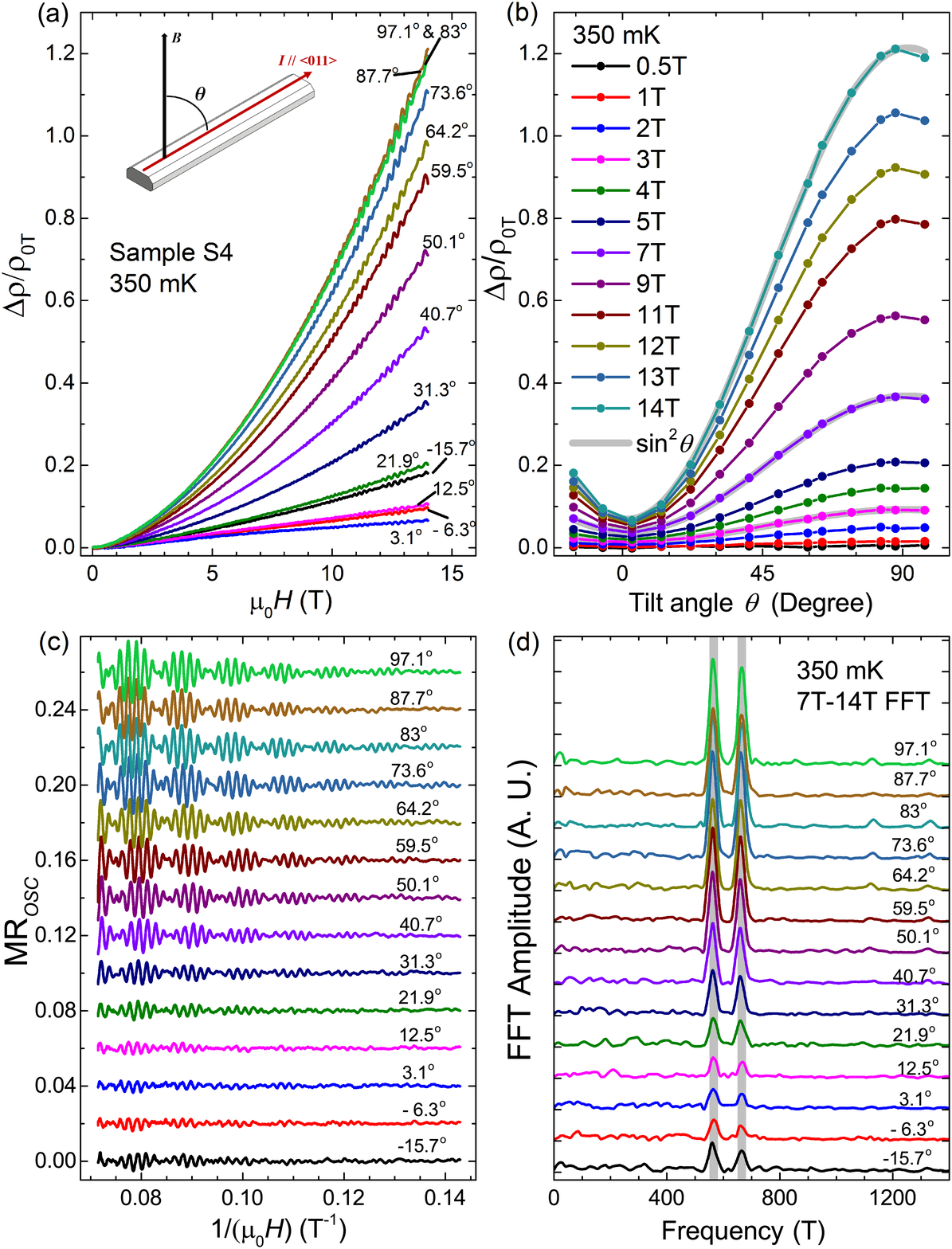}
\caption{\label{fig-MRA}(Color online) (a) Angular dependent MR of CoSi at 350 mK, the inset shows the measurement configuration. (b) shows the MR at given fields extracted from (a), points are connected by lines as a guide to the eyes. Fittings for 3 T, 7 T, and 14 T by sin$^{2}$($\theta$) are superimposed as dark grey lines. The oscillating parts of MR are in (c), and corresponding FFT spectra in (d), all stacked for clarity. In (d), those two vertical grey bars are centered at 564 T and 665 T, respectively. The width of the bars is 30 T.}
\end{figure}
With the aim of probing the bulk Fermi surface properties of CoSi, angular and temperature dependent magnetoresistivity (MR) measurements were carried out on sample S4. Thanks to the thin rod shape of the crystal, the mixing of MR and Hall signals is minimal. Fig.\ref{fig-MRA} (a) shows the isothermal MR curves covering over 90$^{\circ}$, the measurement configuration is illustrated as the inset, both \textit{B} and \textit{I} are parallel to the $\langle011\rangle$ direction at $\theta$ = 0$^{\circ}$. The MR reaches 120\% at 14 T, 350 mK, and $\theta\sim$90$^{\circ}$, about 2 times larger than reported values for CoSi\cite{JMMM.177.1371}, although it is at least one order of magnitude smaller than the huge transverse MR reported for other Dirac and Weyl semimetals\cite{Xu16}. The MR versus tilt angle at selected magnetic fields is plotted in Fig.\ref{fig-MRA} (b), which is symmetric with respect to 0$^{\circ}$ and 90$^{\circ}$ as expected. The data could be tracked by a sin$^{2}$($\theta$) function, suggesting that the MR is proportional to a simple power law of H, with the power close to 2, and is mainly orbital. Besides, the mobility of associated charge carriers should not have a significant anisotropy. The SdH is already apparent as small wiggles on those MR curves in Fig.\ref{fig-MRA} (a). Since the amplitude of the oscillation is less than 2\% of the total resistivity, it is feasible to keep the analysis of SdH on MR only. Fig.\ref{fig-MRA} (c) shows the oscillating part of MR (MR$_{OSC}$) versus the reciprocal of magnetic field after a smooth background substraction by a forth order polynomial. A beating pattern is evident for most of the curves, while the position of nodes of the envelop line has a weak evolution with angle. By fast Fouier transform (FFT) from 7 to 14 T, two prominent fundamental frequencies are identified on the spectra for all the angles stacked in Fig.\ref{fig-MRA} (d), their 1st harmonics are also visible when the oscillation amplitude is strong around $\theta$ = 90$^{\circ}$. By Gaussian peak fit, F1 = 563.9 T and F2 = 664.6T at $\theta\sim$ 87.7$^{\circ}$, and F1 = 564.5 T and F2 = 664.1T at $\theta\sim$ 3.1$^{\circ}$, respectively. The standard errors of fitting are all less than 1 T, while the full width at half maximum (FWHM) is less than 30 T. If the FWHM was taken as the upper limit of frequency shift, then for both F1 and F2 the changes are less than 5\% when \textit{B} rotates from parallel to perpendicular to \textit{I}.

Since the oscillation frequency \emph{F} is directly proportional to the extremal cross-sectional area of a Fermi surface ($A_{F}$) by the Onsager relation\cite{Shoenberg} $F = (\hbar/2\pi e)A_{F}$, here the non-dispersive angular dependence of oscillation frequency F1 and F2 with respect to magnetic field direction indicates that the underlying bulk Fermi surfaces are close to spherical. From recent band structure calculations\cite{PRL.119.206402,JPCond2018}, it is shown that those bands crossing the Fermi level close to the BZ center $\Gamma$ point have their node above the Fermi energy (E$_{F}$), while the Fermi wavevector (k$_{F}$) is different along $\Gamma$-X, $\Gamma$-M, and $\Gamma$-R, which are directions from $\Gamma$ to the face center, the edge center, and the corner of BZ, respectively. At the same time, for those bands crossing close to the R point, their node is $\sim$ 200 meV below E$_{F}$. Without inclusion of SOC, there are two adjacent bands crossing the Fermi level along R-$\Gamma$, one along R-M, and one along R-X, the k$_{F}$ is similar along the three aforementioned directions. Thus there exist two electron Fermi Surfaces (FSs) close in size centered at R, from which we infer the observed two oscillation frequencies are originated. For F1 = 564 T and F2 = 665 T, the corresponding $A_{F}$1 = 5.39$\times$10$^{-2}$\AA$^{-2}$ and $A_{F}$2 = 6.34$\times$10$^{-2}$\AA$^{-2}$, respectively. The averaged value is compatible with that obtained from E$_{F}$ intensity mapping at R point in recent ARPES experiments\cite{PhysRevLett.122.076402,Rao2019}. Here, quantum oscillation shows the advantage of resolving two frequencies. However, we have not yet detected any oscillation from the hole FSs at $\Gamma$.

\begin{figure}[htbp]
\includegraphics[width=0.95\columnwidth]{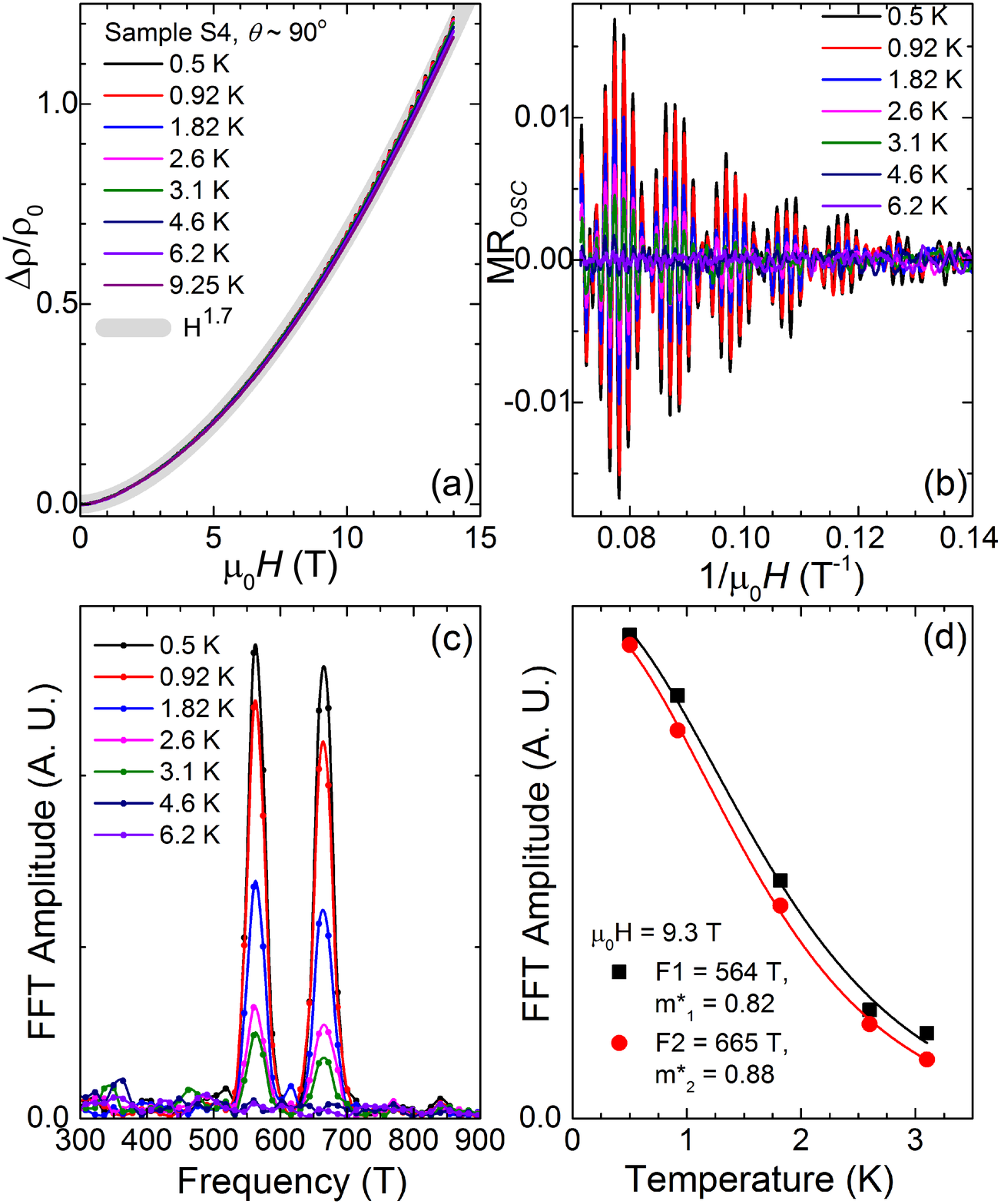}
\caption{\label{fig-MRT}(Color online) (a) Temperature dependent MRs at $\theta\sim$ 90$^{\circ}$, a wide grey curve of MR $\sim$ H$^{1.7}$ is set as background. (b) and (c) are the oscillating parts of MRs and corresponding FFT spectra, respectively. (d) shows fittings of FFT peak amplitude by the thermal damping factor R$_{T}$, an average B of the FFT range, B = 9.3 T is used.}
\end{figure}
For the observed oscillations, more information about charge carriers at the FS could be extracted utilizing the Lifshitz-Kosevitch (LK) formula\cite{Shoenberg,Murakawa1490}, considering summations over different oscillation frequencies and their harmonics,
\begin{align}
\begin{split}
&\textit{MR}_{OSC}\sim\\ &\sum_{i}\sum_{p}\frac{5}{2}\sqrt{\frac{B}{2pF_{i}}}R_{T}^{p}R_{D}^{p}R_{s}^{p} \cos\{2\pi p[\frac{F_{i}}{B}+\frac{1}{2}] + \delta_{i} + \phi_{i}\},\\
\end{split}
\end{align}
in which $R_{T}^{p}=\alpha p m^{*} (T/B)/\sinh[\alpha p m^{*} (T/B)]$, $R_{D}^{p}=\exp[-\alpha p m^{*}(T_{D}/B)]$, and $R_{s}^{p}=\cos(p\pi g m^{*}/2)$ are the thermal damping factor, the Dingle damping factor, and the spin factor, respectively. $\alpha=2\pi^{2}k_{B}m_{e}/e\hbar$ is a prefactor constructed by the Boltzmann constant $k_{B}$, the bare electron mass $m_{e}$, the electron charge \textit{e}, and the reduced Plank constant $\hbar$. For the additional phase shift of the cosine function, $\delta$ is the dimensional correction, and $\phi$ is often denoted as the Berry phase. The effective mass $m^{*}$ at a given field direction can be obtained by the temperature dependence of oscillation amplitude through R$_{T}$. Fig.\ref{fig-MRT} shows the analysis of MR data at $\theta\sim$ 90$^{\circ}$ up to $\sim$ 9 K. The effective masses for previously determined F1 and F2 are m$_{1}^{*}$ = 0.82 and m$_{2}^{*}$ = 0.88, respectively. While in early experimental work, the effective mass of electrons was estimated to be 2\cite{PhysRev.134.A774}. By assuming a circular cross-section of FS, the Fermi wave vectors are estimated as $k_{F1}$ = $1.31\times 10^{-1}$ $\AA^{-1}$, and $k_{F2}$ = $1.42\times10^{-1}$ $\AA^{-1}$, and the Fermi velocities ($v_{F}=\hbar k_{F}/m^{*}$) are $v_{F1}=1.85\times 10^{5} m\cdot s^{-1}$ and $v_{F2}=1.87\times 10^{5} m\cdot s^{-1}$, respectively. As a comparison, $v_{F}$ value of Cd$_{3}$As$_{2}$ from quantum oscillation experiments is around $1\times 10^{6} m\cdot s^{-1}$ \cite{Cd3As2_MR,PhysRevLett.113.246402}.

\begin{figure}[htbp]
\includegraphics[width=0.95\columnwidth]{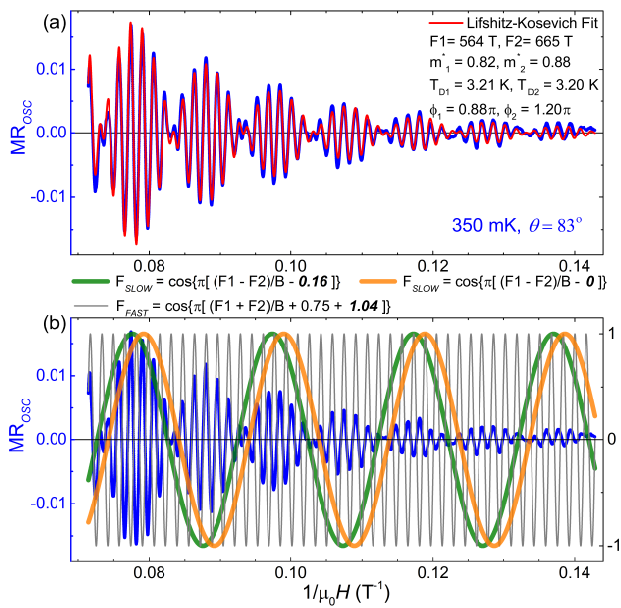}
\caption{\label{fig-LK}(Color online) (a) LK formula (see text) fit of one $MR_{OSC}$ curve with large oscillation amplitude. Fitting parameters are listed in legend. (b) illustrates how the beating pattern is constructed by the product of two cosine functions with periods determined by the difference and summation of two oscillation frequencies, respectively. Note the phase shifts.
}
\end{figure}
Since there are two nearby oscillation frequencies, the extraction of Dingle temperature T$_{D}$ is performed by a direct fitting of $\textit{MR}_{OSC}$ to the LK formula with known F and m$^{*}$. In addition, the Land\'{e} g-factor in $R_{s}$ takes simply the free electron value of 2.0023, and the dimensional phase factor $\delta$ for both F1 and F2 takes the value of (-1/4)$\pi$ for three dimensional electron FS with maximum cross section. By only including fundamental harmonics, a good fit is achieved, as shown in Fig.\ref{fig-LK} (a) for $\theta$ = 83$^{\circ}$, while the residual mainly contains 2nd harmonics. Within fitting error the obtained $T_{D}$ for both F1 and F2 is 3.2 K. The quantum scattering time $\tau_{q}=\hbar/(2\pi k_{B}T_{D})$ for electrons is then computed to be $\sim3.8\times 10^{-13} s$. By taking the values of electron mobility and effective mass from Ref.\cite{PhysRev.134.A774}, $\mu_{e} = $ 800 $cm^{2} V^{-1}s^{-1}$ and $m^{*}=2$, the transport scattering time $\tau_{t}=\mu m^{*}/e$ = $9.1\times 10^{-13} s$. The small difference between $\tau_{t}$ and $\tau_{q}$ could be one reason for the relatively small MR in CoSi when compared to other topological semimetals. As in Cd3As2, $\tau_{t}$ is 4 orders larger than $\tau_{q}$ because of the suppression of backscattering, which is lifted in magnetic field and leads to huge positive MR\cite{Cd3As2_MR}.

The importance of phase factor in quantum oscillation associated with the symmetry of FS has recently been further stressed\cite{PhysRevX.8.011027}. In the above LK fitting, the Berry phase is free parameter and non-zero values within [0,2$\pi$] for both F1 and F2 are obtained. However, we treat them as nominal value and make our interpretation in another route without Landau level index plot. Since the two observed frequencies are close with similar effective masses and Dingle temperatures, a clear beating pattern is present. Then without the field dependent damping factors, the envelop line is given by a 'slow' cosine function $\cos[\pi(F1-F2)/B+(\delta_{1}-\delta_{2})/2 + (\phi_{1}-\phi_{2})/2]$. Given that the dimensional correction terms $\delta_{i}$ are the same for F1 and F2, the position of nodes of the envelop line is dependent on $(\phi_{1}-\phi_{2})/2$. It is illustrated in Fig.\ref{fig-LK} (b) that a finite value of phase shift in $\cos[\pi(F1-F2)/B]$ is needed to correctly track the beating nodes, thus the difference of Berry phase terms for F1 and F2 is non-zero at this field direction, here the absolute value $\sim$0.32$\pi$. As a deduction, at least one of the terms should have a non-zero value, possibly due to chirality of the associated Fermions. For our current angular dependent $MR_{OSC}$ of CoSi, the node positions of the envelop line with respect to field direction have a weak but observable evolution, meanwhile the oscillation frequencies are almost constant, then following the above analysis, without invoking other trivial phase factors, the difference of the two Berry phase terms will show an angular dependence, this could serve as a constraint when analyzing how the individual Berry phase term associated with a FS changes with respect to different k-space directions.

In summary, decent quality CoSi single crystals have been grown with the aid of tin flux. QO originated from the bulk electron FSs at the R point of BZ are observed in magnetoresistivity, and the additional phase in QO arising from electron orbit geometry varies with field direction with non-zero values present. However, a full characterization of the bulk electronic structure of CoSi is still at the beginning stage. Further studies are needed to probe the electronic structure at $\Gamma$, due to the heavier effective mass, we planned to carry out transport and magnetization measurements at lower temperature and up to higher magnetic field.

\textit{Note added}  During preparation of this manuscript, we become aware a preprint by Xu \textit{et. al.}\cite{2019arXiv190400630X}. The authors used Te flux for crystal growth and reached a better RRR for CoSi. QO was observed and yields similar frequencies as ours.

We thank H. M. Weng, Y J. Sun and T. Qian for helpful discussions. This work is supported by the Natural Science Foundation of China with grant No. 11874399, and the National Key Research Program of China with Grant No. 2016YFA0300602.

\bibliography{References}

\begin{thebibliography}{29}%
\makeatletter
\providecommand \@ifxundefined [1]{%
 \@ifx{#1\undefined}
}%
\providecommand \@ifnum [1]{%
 \ifnum #1\expandafter \@firstoftwo
 \else \expandafter \@secondoftwo
 \fi
}%
\providecommand \@ifx [1]{%
 \ifx #1\expandafter \@firstoftwo
 \else \expandafter \@secondoftwo
 \fi
}%
\providecommand \natexlab [1]{#1}%
\providecommand \enquote  [1]{``#1''}%
\providecommand \bibnamefont  [1]{#1}%
\providecommand \bibfnamefont [1]{#1}%
\providecommand \citenamefont [1]{#1}%
\providecommand \href@noop [0]{\@secondoftwo}%
\providecommand \href [0]{\begingroup \@sanitize@url \@href}%
\providecommand \@href[1]{\@@startlink{#1}\@@href}%
\providecommand \@@href[1]{\endgroup#1\@@endlink}%
\providecommand \@sanitize@url [0]{\catcode `\\12\catcode `\$12\catcode
  `\&12\catcode `\#12\catcode `\^12\catcode `\_12\catcode `\%12\relax}%
\providecommand \@@startlink[1]{}%
\providecommand \@@endlink[0]{}%
\providecommand \url  [0]{\begingroup\@sanitize@url \@url }%
\providecommand \@url [1]{\endgroup\@href {#1}{\urlprefix }}%
\providecommand \urlprefix  [0]{URL }%
\providecommand \Eprint [0]{\href }%
\providecommand \doibase [0]{http://dx.doi.org/}%
\providecommand \selectlanguage [0]{\@gobble}%
\providecommand \bibinfo  [0]{\@secondoftwo}%
\providecommand \bibfield  [0]{\@secondoftwo}%
\providecommand \translation [1]{[#1]}%
\providecommand \BibitemOpen [0]{}%
\providecommand \bibitemStop [0]{}%
\providecommand \bibitemNoStop [0]{.\EOS\space}%
\providecommand \EOS [0]{\spacefactor3000\relax}%
\providecommand \BibitemShut  [1]{\csname bibitem#1\endcsname}%
\let\auto@bib@innerbib\@empty
\bibitem [{\citenamefont {Sakai}\ \emph {et~al.}(2007)\citenamefont {Sakai},
  \citenamefont {Ishii}, \citenamefont {Onose}, \citenamefont {Tomioka},
  \citenamefont {Yotsuhashi}, \citenamefont {Adachi}, \citenamefont {Nagaosa},\
  and\ \citenamefont {Tokura}}]{JPSJ.76.093601}%
  \BibitemOpen
  \bibfield  {author} {\bibinfo {author} {\bibfnamefont {A.}~\bibnamefont
  {Sakai}}, \bibinfo {author} {\bibfnamefont {F.}~\bibnamefont {Ishii}},
  \bibinfo {author} {\bibfnamefont {Y.}~\bibnamefont {Onose}}, \bibinfo
  {author} {\bibfnamefont {Y.}~\bibnamefont {Tomioka}}, \bibinfo {author}
  {\bibfnamefont {S.}~\bibnamefont {Yotsuhashi}}, \bibinfo {author}
  {\bibfnamefont {H.}~\bibnamefont {Adachi}}, \bibinfo {author} {\bibfnamefont
  {N.}~\bibnamefont {Nagaosa}}, \ and\ \bibinfo {author} {\bibfnamefont
  {Y.}~\bibnamefont {Tokura}},\ }\href {\doibase 10.1143/JPSJ.76.093601}
  {\bibfield  {journal} {\bibinfo  {journal} {JPSJ}\ }\textbf {\bibinfo
  {volume} {76}},\ \bibinfo {pages} {093601} (\bibinfo {year}
  {2007})}\BibitemShut {NoStop}%
\bibitem [{\citenamefont {Longhin}\ \emph {et~al.}(2017)\citenamefont
  {Longhin}, \citenamefont {Rizza}, \citenamefont {Viennois},\ and\
  \citenamefont {Papet}}]{Inte.Metal.88.46}%
  \BibitemOpen
  \bibfield  {author} {\bibinfo {author} {\bibfnamefont {M.}~\bibnamefont
  {Longhin}}, \bibinfo {author} {\bibfnamefont {M.}~\bibnamefont {Rizza}},
  \bibinfo {author} {\bibfnamefont {R.}~\bibnamefont {Viennois}}, \ and\
  \bibinfo {author} {\bibfnamefont {P.}~\bibnamefont {Papet}},\ }\href
  {\doibase https://doi.org/10.1016/j.intermet.2017.04.014} {\bibfield
  {journal} {\bibinfo  {journal} {Intermetallics}\ }\textbf {\bibinfo {volume}
  {88}},\ \bibinfo {pages} {46 } (\bibinfo {year} {2017})}\BibitemShut
  {NoStop}%
\bibitem [{\citenamefont {Bradlyn}\ \emph {et~al.}(2016)\citenamefont
  {Bradlyn}, \citenamefont {Cano}, \citenamefont {Wang}, \citenamefont
  {Vergniory}, \citenamefont {Felser}, \citenamefont {Cava},\ and\
  \citenamefont {Bernevig}}]{TPSM_Science}%
  \BibitemOpen
  \bibfield  {author} {\bibinfo {author} {\bibfnamefont {B.}~\bibnamefont
  {Bradlyn}}, \bibinfo {author} {\bibfnamefont {J.}~\bibnamefont {Cano}},
  \bibinfo {author} {\bibfnamefont {Z.}~\bibnamefont {Wang}}, \bibinfo {author}
  {\bibfnamefont {M.~G.}\ \bibnamefont {Vergniory}}, \bibinfo {author}
  {\bibfnamefont {C.}~\bibnamefont {Felser}}, \bibinfo {author} {\bibfnamefont
  {R.~J.}\ \bibnamefont {Cava}}, \ and\ \bibinfo {author} {\bibfnamefont
  {B.~A.}\ \bibnamefont {Bernevig}},\ }\href {\doibase 10.1126/science.aaf5037}
  {\bibfield  {journal} {\bibinfo  {journal} {Science}\ }\textbf {\bibinfo
  {volume} {353}},\ \bibinfo {pages} {aaf5037} (\bibinfo {year}
  {2016})}\BibitemShut {NoStop}%
\bibitem [{\citenamefont {Tang}\ \emph {et~al.}(2017)\citenamefont {Tang},
  \citenamefont {Zhou},\ and\ \citenamefont {Zhang}}]{PRL.119.206402}%
  \BibitemOpen
  \bibfield  {author} {\bibinfo {author} {\bibfnamefont {P.}~\bibnamefont
  {Tang}}, \bibinfo {author} {\bibfnamefont {Q.}~\bibnamefont {Zhou}}, \ and\
  \bibinfo {author} {\bibfnamefont {S.-C.}\ \bibnamefont {Zhang}},\ }\href
  {\doibase 10.1103/PhysRevLett.119.206402} {\bibfield  {journal} {\bibinfo
  {journal} {Phys. Rev. Lett.}\ }\textbf {\bibinfo {volume} {119}},\ \bibinfo
  {pages} {206402} (\bibinfo {year} {2017})},\ \bibinfo {note} {and its
  supplemenatry material}\BibitemShut {NoStop}%
\bibitem [{\citenamefont {Chang}\ \emph {et~al.}(2017)\citenamefont {Chang},
  \citenamefont {Xu}, \citenamefont {Wieder}, \citenamefont {Sanchez},
  \citenamefont {Huang}, \citenamefont {Belopolski}, \citenamefont {Chang},
  \citenamefont {Zhang}, \citenamefont {Bansil}, \citenamefont {Lin},\ and\
  \citenamefont {Hasan}}]{PRL.119.206401}%
  \BibitemOpen
  \bibfield  {author} {\bibinfo {author} {\bibfnamefont {G.}~\bibnamefont
  {Chang}}, \bibinfo {author} {\bibfnamefont {S.-Y.}\ \bibnamefont {Xu}},
  \bibinfo {author} {\bibfnamefont {B.~J.}\ \bibnamefont {Wieder}}, \bibinfo
  {author} {\bibfnamefont {D.~S.}\ \bibnamefont {Sanchez}}, \bibinfo {author}
  {\bibfnamefont {S.-M.}\ \bibnamefont {Huang}}, \bibinfo {author}
  {\bibfnamefont {I.}~\bibnamefont {Belopolski}}, \bibinfo {author}
  {\bibfnamefont {T.-R.}\ \bibnamefont {Chang}}, \bibinfo {author}
  {\bibfnamefont {S.}~\bibnamefont {Zhang}}, \bibinfo {author} {\bibfnamefont
  {A.}~\bibnamefont {Bansil}}, \bibinfo {author} {\bibfnamefont
  {H.}~\bibnamefont {Lin}}, \ and\ \bibinfo {author} {\bibfnamefont {M.~Z.}\
  \bibnamefont {Hasan}},\ }\href {\doibase 10.1103/PhysRevLett.119.206401}
  {\bibfield  {journal} {\bibinfo  {journal} {Phys. Rev. Lett.}\ }\textbf
  {\bibinfo {volume} {119}},\ \bibinfo {pages} {206401} (\bibinfo {year}
  {2017})}\BibitemShut {NoStop}%
\bibitem [{\citenamefont {Zhang}\ \emph {et~al.}(2018)\citenamefont {Zhang},
  \citenamefont {Song}, \citenamefont {Alexandradinata}, \citenamefont {Weng},
  \citenamefont {Fang}, \citenamefont {Lu},\ and\ \citenamefont
  {Fang}}]{PRL.120.016401}%
  \BibitemOpen
  \bibfield  {author} {\bibinfo {author} {\bibfnamefont {T.}~\bibnamefont
  {Zhang}}, \bibinfo {author} {\bibfnamefont {Z.}~\bibnamefont {Song}},
  \bibinfo {author} {\bibfnamefont {A.}~\bibnamefont {Alexandradinata}},
  \bibinfo {author} {\bibfnamefont {H.}~\bibnamefont {Weng}}, \bibinfo {author}
  {\bibfnamefont {C.}~\bibnamefont {Fang}}, \bibinfo {author} {\bibfnamefont
  {L.}~\bibnamefont {Lu}}, \ and\ \bibinfo {author} {\bibfnamefont
  {Z.}~\bibnamefont {Fang}},\ }\href {\doibase 10.1103/PhysRevLett.120.016401}
  {\bibfield  {journal} {\bibinfo  {journal} {Phys. Rev. Lett.}\ }\textbf
  {\bibinfo {volume} {120}},\ \bibinfo {pages} {016401} (\bibinfo {year}
  {2018})}\BibitemShut {NoStop}%
\bibitem [{\citenamefont {Takane}\ \emph {et~al.}(2019)\citenamefont {Takane},
  \citenamefont {Wang}, \citenamefont {Souma}, \citenamefont {Nakayama},
  \citenamefont {Nakamura}, \citenamefont {Oinuma}, \citenamefont {Nakata},
  \citenamefont {Iwasawa}, \citenamefont {Cacho}, \citenamefont {Kim},
  \citenamefont {Horiba}, \citenamefont {Kumigashira}, \citenamefont
  {Takahashi}, \citenamefont {Ando},\ and\ \citenamefont
  {Sato}}]{PhysRevLett.122.076402}%
  \BibitemOpen
  \bibfield  {author} {\bibinfo {author} {\bibfnamefont {D.}~\bibnamefont
  {Takane}}, \bibinfo {author} {\bibfnamefont {Z.}~\bibnamefont {Wang}},
  \bibinfo {author} {\bibfnamefont {S.}~\bibnamefont {Souma}}, \bibinfo
  {author} {\bibfnamefont {K.}~\bibnamefont {Nakayama}}, \bibinfo {author}
  {\bibfnamefont {T.}~\bibnamefont {Nakamura}}, \bibinfo {author}
  {\bibfnamefont {H.}~\bibnamefont {Oinuma}}, \bibinfo {author} {\bibfnamefont
  {Y.}~\bibnamefont {Nakata}}, \bibinfo {author} {\bibfnamefont
  {H.}~\bibnamefont {Iwasawa}}, \bibinfo {author} {\bibfnamefont
  {C.}~\bibnamefont {Cacho}}, \bibinfo {author} {\bibfnamefont
  {T.}~\bibnamefont {Kim}}, \bibinfo {author} {\bibfnamefont {K.}~\bibnamefont
  {Horiba}}, \bibinfo {author} {\bibfnamefont {H.}~\bibnamefont {Kumigashira}},
  \bibinfo {author} {\bibfnamefont {T.}~\bibnamefont {Takahashi}}, \bibinfo
  {author} {\bibfnamefont {Y.}~\bibnamefont {Ando}}, \ and\ \bibinfo {author}
  {\bibfnamefont {T.}~\bibnamefont {Sato}},\ }\href {\doibase
  10.1103/PhysRevLett.122.076402} {\bibfield  {journal} {\bibinfo  {journal}
  {Phys. Rev. Lett.}\ }\textbf {\bibinfo {volume} {122}},\ \bibinfo {pages}
  {076402} (\bibinfo {year} {2019})}\BibitemShut {NoStop}%
\bibitem [{\citenamefont {Rao}\ \emph {et~al.}(2019)\citenamefont {Rao},
  \citenamefont {Li}, \citenamefont {Zhang}, \citenamefont {Tian},
  \citenamefont {Li}, \citenamefont {Fu} \emph {et~al.}}]{Rao2019}%
  \BibitemOpen
  \bibfield  {author} {\bibinfo {author} {\bibfnamefont {Z.~C.}\ \bibnamefont
  {Rao}}, \bibinfo {author} {\bibfnamefont {H.}~\bibnamefont {Li}}, \bibinfo
  {author} {\bibfnamefont {T.~T.}\ \bibnamefont {Zhang}}, \bibinfo {author}
  {\bibfnamefont {S.~J.}\ \bibnamefont {Tian}}, \bibinfo {author}
  {\bibfnamefont {C.~H.}\ \bibnamefont {Li}}, \bibinfo {author} {\bibfnamefont
  {B.~B.}\ \bibnamefont {Fu}},  \emph {et~al.},\ }\href {\doibase
  10.1038/s41586-019-1031-8} {\bibfield  {journal} {\bibinfo  {journal}
  {Nature}\ }\textbf {\bibinfo {volume} {567}},\ \bibinfo {pages} {496}
  (\bibinfo {year} {2019})}\BibitemShut {NoStop}%
\bibitem [{\citenamefont {Sanchez}\ \emph {et~al.}(2019)\citenamefont
  {Sanchez}, \citenamefont {Belopolski}, \citenamefont {Cochran}, \citenamefont
  {Xu}, \citenamefont {Yin}, \citenamefont {Chang}, \citenamefont {Xie} \emph
  {et~al.}}]{Sanchez2019}%
  \BibitemOpen
  \bibfield  {author} {\bibinfo {author} {\bibfnamefont {D.~S.}\ \bibnamefont
  {Sanchez}}, \bibinfo {author} {\bibfnamefont {I.}~\bibnamefont {Belopolski}},
  \bibinfo {author} {\bibfnamefont {T.~A.}\ \bibnamefont {Cochran}}, \bibinfo
  {author} {\bibfnamefont {X.}~\bibnamefont {Xu}}, \bibinfo {author}
  {\bibfnamefont {J.-X.}\ \bibnamefont {Yin}}, \bibinfo {author} {\bibfnamefont
  {G.}~\bibnamefont {Chang}}, \bibinfo {author} {\bibfnamefont
  {W.}~\bibnamefont {Xie}},  \emph {et~al.},\ }\href {\doibase
  10.1038/s41586-019-1037-2} {\bibfield  {journal} {\bibinfo  {journal}
  {Nature}\ }\textbf {\bibinfo {volume} {567}},\ \bibinfo {pages} {500}
  (\bibinfo {year} {2019})}\BibitemShut {NoStop}%
\bibitem [{\citenamefont {Ishida}\ \emph {et~al.}(1991)\citenamefont {Ishida},
  \citenamefont {Nishizawa},\ and\ \citenamefont {Schlesinger}}]{Co-Si_phase}%
  \BibitemOpen
  \bibfield  {author} {\bibinfo {author} {\bibfnamefont {K.}~\bibnamefont
  {Ishida}}, \bibinfo {author} {\bibfnamefont {T.}~\bibnamefont {Nishizawa}}, \
  and\ \bibinfo {author} {\bibfnamefont {M.~E.}\ \bibnamefont {Schlesinger}},\
  }\href {\doibase 10.1007/BF02645074} {\bibfield  {journal} {\bibinfo
  {journal} {Journal of phase equilibria}\ }\textbf {\bibinfo {volume} {12}},\
  \bibinfo {pages} {578} (\bibinfo {year} {1991})}\BibitemShut {NoStop}%
\bibitem [{SHE()}]{SHELX}%
  \BibitemOpen
  \href {http://shelx.uni-goettingen.de} {}\bibinfo {note}
  {Http://shelx.uni-goettingen.de}\BibitemShut {NoStop}%
\bibitem [{\citenamefont {Stishov}\ \emph {et~al.}(2012)\citenamefont
  {Stishov}, \citenamefont {Petrova}, \citenamefont {Sidorov},\ and\
  \citenamefont {Menzel}}]{PRB.86.064433}%
  \BibitemOpen
  \bibfield  {author} {\bibinfo {author} {\bibfnamefont {S.~M.}\ \bibnamefont
  {Stishov}}, \bibinfo {author} {\bibfnamefont {A.~E.}\ \bibnamefont
  {Petrova}}, \bibinfo {author} {\bibfnamefont {V.~A.}\ \bibnamefont
  {Sidorov}}, \ and\ \bibinfo {author} {\bibfnamefont {D.}~\bibnamefont
  {Menzel}},\ }\href {\doibase 10.1103/PhysRevB.86.064433} {\bibfield
  {journal} {\bibinfo  {journal} {Phys. Rev. B}\ }\textbf {\bibinfo {volume}
  {86}},\ \bibinfo {pages} {064433} (\bibinfo {year} {2012})}\BibitemShut
  {NoStop}%
\bibitem [{\citenamefont {Amamou}\ \emph {et~al.}(1972)\citenamefont {Amamou},
  \citenamefont {Bach}, \citenamefont {Gautier}, \citenamefont {Robert},\ and\
  \citenamefont {Castaing}}]{HeatCap72}%
  \BibitemOpen
  \bibfield  {author} {\bibinfo {author} {\bibfnamefont {A.}~\bibnamefont
  {Amamou}}, \bibinfo {author} {\bibfnamefont {P.}~\bibnamefont {Bach}},
  \bibinfo {author} {\bibfnamefont {F.}~\bibnamefont {Gautier}}, \bibinfo
  {author} {\bibfnamefont {C.}~\bibnamefont {Robert}}, \ and\ \bibinfo {author}
  {\bibfnamefont {J.}~\bibnamefont {Castaing}},\ }\href {\doibase
  https://doi.org/10.1016/S0022-3697(72)80465-7} {\bibfield  {journal}
  {\bibinfo  {journal} {J. Phys. Chem. Solids}\ }\textbf {\bibinfo {volume}
  {33}},\ \bibinfo {pages} {1697 } (\bibinfo {year} {1972})}\BibitemShut
  {NoStop}%
\bibitem [{\citenamefont {Narozhnyi}\ and\ \citenamefont
  {Krasnorussky}(2013)}]{Narozhnyi2013}%
  \BibitemOpen
  \bibfield  {author} {\bibinfo {author} {\bibfnamefont {V.~N.}\ \bibnamefont
  {Narozhnyi}}\ and\ \bibinfo {author} {\bibfnamefont {V.~N.}\ \bibnamefont
  {Krasnorussky}},\ }\href {\doibase 10.1134/S106377611305021X} {\bibfield
  {journal} {\bibinfo  {journal} {JETP}\ }\textbf {\bibinfo {volume} {116}},\
  \bibinfo {pages} {780} (\bibinfo {year} {2013})},\ \bibinfo {note} {and
  reference therein}\BibitemShut {NoStop}%
\bibitem [{\citenamefont {{Burkov}}\ \emph {et~al.}(2017)\citenamefont
  {{Burkov}}, \citenamefont {{Novikov}}, \citenamefont {{Zaitsev}},\ and\
  \citenamefont {{Reith}}}]{Semicon}%
  \BibitemOpen
  \bibfield  {author} {\bibinfo {author} {\bibfnamefont {A.~T.}\ \bibnamefont
  {{Burkov}}}, \bibinfo {author} {\bibfnamefont {S.~V.}\ \bibnamefont
  {{Novikov}}}, \bibinfo {author} {\bibfnamefont {V.~K.}\ \bibnamefont
  {{Zaitsev}}}, \ and\ \bibinfo {author} {\bibfnamefont {H.}~\bibnamefont
  {{Reith}}},\ }\href {\doibase 10.1134/S1063782617060094} {\bibfield
  {journal} {\bibinfo  {journal} {Semiconductors}\ }\textbf {\bibinfo {volume}
  {51}},\ \bibinfo {pages} {689} (\bibinfo {year} {2017})}\BibitemShut
  {NoStop}%
\bibitem [{\citenamefont {Petrova}\ \emph {et~al.}(2010)\citenamefont
  {Petrova}, \citenamefont {Krasnorussky}, \citenamefont {Shikov},
  \citenamefont {Yuhasz}, \citenamefont {Lograsso}, \citenamefont {Lashley},\
  and\ \citenamefont {Stishov}}]{PRB.82.155124}%
  \BibitemOpen
  \bibfield  {author} {\bibinfo {author} {\bibfnamefont {A.~E.}\ \bibnamefont
  {Petrova}}, \bibinfo {author} {\bibfnamefont {V.~N.}\ \bibnamefont
  {Krasnorussky}}, \bibinfo {author} {\bibfnamefont {A.~A.}\ \bibnamefont
  {Shikov}}, \bibinfo {author} {\bibfnamefont {W.~M.}\ \bibnamefont {Yuhasz}},
  \bibinfo {author} {\bibfnamefont {T.~A.}\ \bibnamefont {Lograsso}}, \bibinfo
  {author} {\bibfnamefont {J.~C.}\ \bibnamefont {Lashley}}, \ and\ \bibinfo
  {author} {\bibfnamefont {S.~M.}\ \bibnamefont {Stishov}},\ }\href {\doibase
  10.1103/PhysRevB.82.155124} {\bibfield  {journal} {\bibinfo  {journal} {Phys.
  Rev. B}\ }\textbf {\bibinfo {volume} {82}},\ \bibinfo {pages} {155124}
  (\bibinfo {year} {2010})}\BibitemShut {NoStop}%
\bibitem [{\citenamefont {Ohta}\ \emph {et~al.}(1998)\citenamefont {Ohta},
  \citenamefont {Arioka}, \citenamefont {Kulatov}, \citenamefont {Mitsudo},\
  and\ \citenamefont {Motokawa}}]{JMMM.177.1371}%
  \BibitemOpen
  \bibfield  {author} {\bibinfo {author} {\bibfnamefont {H.}~\bibnamefont
  {Ohta}}, \bibinfo {author} {\bibfnamefont {T.}~\bibnamefont {Arioka}},
  \bibinfo {author} {\bibfnamefont {E.}~\bibnamefont {Kulatov}}, \bibinfo
  {author} {\bibfnamefont {S.}~\bibnamefont {Mitsudo}}, \ and\ \bibinfo
  {author} {\bibfnamefont {M.}~\bibnamefont {Motokawa}},\ }\href {\doibase
  https://doi.org/10.1016/S0304-8853(97)00409-5} {\bibfield  {journal}
  {\bibinfo  {journal} {J. Magn. Magn. Mater.}\ }\textbf {\bibinfo {volume}
  {177-181}},\ \bibinfo {pages} {1371 } (\bibinfo {year} {1998})}\BibitemShut
  {NoStop}%
\bibitem [{\citenamefont {Ou-Yang}\ \emph {et~al.}(2017)\citenamefont
  {Ou-Yang}, \citenamefont {Shu},\ and\ \citenamefont {Fuh}}]{EPL.120.17002}%
  \BibitemOpen
  \bibfield  {author} {\bibinfo {author} {\bibfnamefont {T.~Y.}\ \bibnamefont
  {Ou-Yang}}, \bibinfo {author} {\bibfnamefont {G.~J.}\ \bibnamefont {Shu}}, \
  and\ \bibinfo {author} {\bibfnamefont {H.~R.}\ \bibnamefont {Fuh}},\ }\href
  {\doibase 10.1209/0295-5075/120/17002} {\bibfield  {journal} {\bibinfo
  {journal} {{EPL} (Europhysics Letters)}\ }\textbf {\bibinfo {volume} {120}},\
  \bibinfo {pages} {17002} (\bibinfo {year} {2017})}\BibitemShut {NoStop}%
\bibitem [{\citenamefont {van~der Marel}\ \emph {et~al.}(1998)\citenamefont
  {van~der Marel}, \citenamefont {Damascelli}, \citenamefont {Schulte},\ and\
  \citenamefont {Menovsky}}]{PhysicaB.244.138}%
  \BibitemOpen
  \bibfield  {author} {\bibinfo {author} {\bibfnamefont {D.}~\bibnamefont
  {van~der Marel}}, \bibinfo {author} {\bibfnamefont {A.}~\bibnamefont
  {Damascelli}}, \bibinfo {author} {\bibfnamefont {K.}~\bibnamefont {Schulte}},
  \ and\ \bibinfo {author} {\bibfnamefont {A.}~\bibnamefont {Menovsky}},\
  }\href {\doibase https://doi.org/10.1016/S0921-4526(97)00476-6} {\bibfield
  {journal} {\bibinfo  {journal} {Physica B: Condens. Matt.}\ }\textbf
  {\bibinfo {volume} {244}},\ \bibinfo {pages} {138 } (\bibinfo {year}
  {1998})}\BibitemShut {NoStop}%
\bibitem [{\citenamefont {Liang}\ \emph {et~al.}(2015)\citenamefont {Liang},
  \citenamefont {Gibson}, \citenamefont {Ali}, \citenamefont {Liu},
  \citenamefont {Cava},\ and\ \citenamefont {Ong}}]{Cd3As2_MR}%
  \BibitemOpen
  \bibfield  {author} {\bibinfo {author} {\bibfnamefont {T.}~\bibnamefont
  {Liang}}, \bibinfo {author} {\bibfnamefont {Q.}~\bibnamefont {Gibson}},
  \bibinfo {author} {\bibfnamefont {M.~N.}\ \bibnamefont {Ali}}, \bibinfo
  {author} {\bibfnamefont {M.}~\bibnamefont {Liu}}, \bibinfo {author}
  {\bibfnamefont {R.~J.}\ \bibnamefont {Cava}}, \ and\ \bibinfo {author}
  {\bibfnamefont {N.~P.}\ \bibnamefont {Ong}},\ }\href@noop {} {\bibfield
  {journal} {\bibinfo  {journal} {Nat. Mater.}\ }\textbf {\bibinfo {volume}
  {14}},\ \bibinfo {pages} {280} (\bibinfo {year} {2015})}\BibitemShut
  {NoStop}%
\bibitem [{\citenamefont {Cvijovi{\'{c}}}(2011)}]{Cvijovi2011}%
  \BibitemOpen
  \bibfield  {author} {\bibinfo {author} {\bibfnamefont {D.}~\bibnamefont
  {Cvijovi{\'{c}}}},\ }\href {\doibase 10.1007/s11232-011-0003-4} {\bibfield
  {journal} {\bibinfo  {journal} {Theo. Math. Phys.}\ }\textbf {\bibinfo
  {volume} {166}},\ \bibinfo {pages} {37} (\bibinfo {year} {2011})}\BibitemShut
  {NoStop}%
\bibitem [{\citenamefont {Xu}\ and\ \citenamefont {Jia}(2016)}]{Xu16}%
  \BibitemOpen
  \bibfield  {author} {\bibinfo {author} {\bibfnamefont {X.-T.}\ \bibnamefont
  {Xu}}\ and\ \bibinfo {author} {\bibfnamefont {S.}~\bibnamefont {Jia}},\
  }\href {\doibase 10.1088/1674-1056/25/11/117204} {\bibfield  {journal}
  {\bibinfo  {journal} {Chin. Phys. B}\ }\textbf {\bibinfo {volume} {25}},\
  \bibinfo {pages} {117204} (\bibinfo {year} {2016})}\BibitemShut {NoStop}%
\bibitem [{\citenamefont {Shoenberg}(1984)}]{Shoenberg}%
  \BibitemOpen
  \bibfield  {author} {\bibinfo {author} {\bibfnamefont {D.}~\bibnamefont
  {Shoenberg}},\ }\href@noop {} {\emph {\bibinfo {title} {Magnetic oscillations
  in metals}}}\ (\bibinfo  {publisher} {Cambridge University, Cambridge,
  England},\ \bibinfo {year} {1984})\BibitemShut {NoStop}%
\bibitem [{\citenamefont {Pshenay-Severin}\ \emph {et~al.}(2018)\citenamefont
  {Pshenay-Severin}, \citenamefont {Ivanov}, \citenamefont {Burkov},\ and\
  \citenamefont {Burkov}}]{JPCond2018}%
  \BibitemOpen
  \bibfield  {author} {\bibinfo {author} {\bibfnamefont {D.~A.}\ \bibnamefont
  {Pshenay-Severin}}, \bibinfo {author} {\bibfnamefont {Y.~V.}\ \bibnamefont
  {Ivanov}}, \bibinfo {author} {\bibfnamefont {A.~A.}\ \bibnamefont {Burkov}},
  \ and\ \bibinfo {author} {\bibfnamefont {A.~T.}\ \bibnamefont {Burkov}},\
  }\href {\doibase 10.1088/1361-648x/aab0ba} {\bibfield  {journal} {\bibinfo
  {journal} {J. Phys.: Condens. Matt.}\ }\textbf {\bibinfo {volume} {30}},\
  \bibinfo {pages} {135501} (\bibinfo {year} {2018})}\BibitemShut {NoStop}%
\bibitem [{\citenamefont {Murakawa}\ \emph {et~al.}(2013)\citenamefont
  {Murakawa}, \citenamefont {Bahramy}, \citenamefont {Tokunaga}, \citenamefont
  {Kohama}, \citenamefont {Bell}, \citenamefont {Kaneko}, \citenamefont
  {Nagaosa}, \citenamefont {Hwang},\ and\ \citenamefont
  {Tokura}}]{Murakawa1490}%
  \BibitemOpen
  \bibfield  {author} {\bibinfo {author} {\bibfnamefont {H.}~\bibnamefont
  {Murakawa}}, \bibinfo {author} {\bibfnamefont {M.~S.}\ \bibnamefont
  {Bahramy}}, \bibinfo {author} {\bibfnamefont {M.}~\bibnamefont {Tokunaga}},
  \bibinfo {author} {\bibfnamefont {Y.}~\bibnamefont {Kohama}}, \bibinfo
  {author} {\bibfnamefont {C.}~\bibnamefont {Bell}}, \bibinfo {author}
  {\bibfnamefont {Y.}~\bibnamefont {Kaneko}}, \bibinfo {author} {\bibfnamefont
  {N.}~\bibnamefont {Nagaosa}}, \bibinfo {author} {\bibfnamefont {H.~Y.}\
  \bibnamefont {Hwang}}, \ and\ \bibinfo {author} {\bibfnamefont
  {Y.}~\bibnamefont {Tokura}},\ }\href {\doibase 10.1126/science.1242247}
  {\bibfield  {journal} {\bibinfo  {journal} {Science}\ }\textbf {\bibinfo
  {volume} {342}},\ \bibinfo {pages} {1490} (\bibinfo {year} {2013})},\
  \bibinfo {note} {and its supplementary materials}\BibitemShut {NoStop}%
\bibitem [{\citenamefont {Asanabe}\ \emph {et~al.}(1964)\citenamefont
  {Asanabe}, \citenamefont {Shinoda},\ and\ \citenamefont
  {Sasaki}}]{PhysRev.134.A774}%
  \BibitemOpen
  \bibfield  {author} {\bibinfo {author} {\bibfnamefont {S.}~\bibnamefont
  {Asanabe}}, \bibinfo {author} {\bibfnamefont {D.}~\bibnamefont {Shinoda}}, \
  and\ \bibinfo {author} {\bibfnamefont {Y.}~\bibnamefont {Sasaki}},\ }\href
  {\doibase 10.1103/PhysRev.134.A774} {\bibfield  {journal} {\bibinfo
  {journal} {Phys. Rev.}\ }\textbf {\bibinfo {volume} {134}},\ \bibinfo {pages}
  {A774} (\bibinfo {year} {1964})}\BibitemShut {NoStop}%
\bibitem [{\citenamefont {He}\ \emph {et~al.}(2014)\citenamefont {He},
  \citenamefont {Hong}, \citenamefont {Dong}, \citenamefont {Pan},
  \citenamefont {Zhang}, \citenamefont {Zhang},\ and\ \citenamefont
  {Li}}]{PhysRevLett.113.246402}%
  \BibitemOpen
  \bibfield  {author} {\bibinfo {author} {\bibfnamefont {L.~P.}\ \bibnamefont
  {He}}, \bibinfo {author} {\bibfnamefont {X.~C.}\ \bibnamefont {Hong}},
  \bibinfo {author} {\bibfnamefont {J.~K.}\ \bibnamefont {Dong}}, \bibinfo
  {author} {\bibfnamefont {J.}~\bibnamefont {Pan}}, \bibinfo {author}
  {\bibfnamefont {Z.}~\bibnamefont {Zhang}}, \bibinfo {author} {\bibfnamefont
  {J.}~\bibnamefont {Zhang}}, \ and\ \bibinfo {author} {\bibfnamefont {S.~Y.}\
  \bibnamefont {Li}},\ }\href {\doibase 10.1103/PhysRevLett.113.246402}
  {\bibfield  {journal} {\bibinfo  {journal} {Phys. Rev. Lett.}\ }\textbf
  {\bibinfo {volume} {113}},\ \bibinfo {pages} {246402} (\bibinfo {year}
  {2014})}\BibitemShut {NoStop}%
\bibitem [{\citenamefont {Alexandradinata}\ \emph {et~al.}(2018)\citenamefont
  {Alexandradinata}, \citenamefont {Wang}, \citenamefont {Duan},\ and\
  \citenamefont {Glazman}}]{PhysRevX.8.011027}%
  \BibitemOpen
  \bibfield  {author} {\bibinfo {author} {\bibfnamefont {A.}~\bibnamefont
  {Alexandradinata}}, \bibinfo {author} {\bibfnamefont {C.}~\bibnamefont
  {Wang}}, \bibinfo {author} {\bibfnamefont {W.}~\bibnamefont {Duan}}, \ and\
  \bibinfo {author} {\bibfnamefont {L.}~\bibnamefont {Glazman}},\ }\href
  {\doibase 10.1103/PhysRevX.8.011027} {\bibfield  {journal} {\bibinfo
  {journal} {Phys. Rev. X}\ }\textbf {\bibinfo {volume} {8}},\ \bibinfo {pages}
  {011027} (\bibinfo {year} {2018})}\BibitemShut {NoStop}%
\bibitem [{\citenamefont {{Xu}}\ \emph {et~al.}(2019)\citenamefont {{Xu}},
  \citenamefont {{Wang}}, \citenamefont {{Cochran}}, \citenamefont {{Sanchez}},
  \citenamefont {{Belopolski}}, \citenamefont {{Wang}}, \citenamefont {{Liu}},
  \citenamefont {{Tien}}, \citenamefont {{Gui}}, \citenamefont {{Xie}},
  \citenamefont {{Zahid Hasan}}, \citenamefont {{Chang}},\ and\ \citenamefont
  {{Jia}}}]{2019arXiv190400630X}%
  \BibitemOpen
  \bibfield  {author} {\bibinfo {author} {\bibfnamefont {X.}~\bibnamefont
  {{Xu}}}, \bibinfo {author} {\bibfnamefont {X.}~\bibnamefont {{Wang}}},
  \bibinfo {author} {\bibfnamefont {T.~A.}\ \bibnamefont {{Cochran}}}, \bibinfo
  {author} {\bibfnamefont {D.~S.}\ \bibnamefont {{Sanchez}}}, \bibinfo {author}
  {\bibfnamefont {I.}~\bibnamefont {{Belopolski}}}, \bibinfo {author}
  {\bibfnamefont {G.}~\bibnamefont {{Wang}}}, \bibinfo {author} {\bibfnamefont
  {Y.}~\bibnamefont {{Liu}}}, \bibinfo {author} {\bibfnamefont {H.-J.}\
  \bibnamefont {{Tien}}}, \bibinfo {author} {\bibfnamefont {X.}~\bibnamefont
  {{Gui}}}, \bibinfo {author} {\bibfnamefont {W.}~\bibnamefont {{Xie}}},
  \bibinfo {author} {\bibfnamefont {M.}~\bibnamefont {{Zahid Hasan}}}, \bibinfo
  {author} {\bibfnamefont {T.-R.}\ \bibnamefont {{Chang}}}, \ and\ \bibinfo
  {author} {\bibfnamefont {S.}~\bibnamefont {{Jia}}},\ }\href
  {https://arxiv.org/abs/1904.00630v1} {\  (\bibinfo {year} {2019})},\ \Eprint
  {http://arxiv.org/abs/1904.00630} {arXiv:1904.00630 [cond-mat.mtrl-sci]}
  \BibitemShut {NoStop}%
\end{thebibliography}%

\end{document}